\documentclass{aa}
\usepackage{txfonts}
\usepackage{natbib,twoopt}
\usepackage[breaklinks=true]{hyperref} 
\hypersetup{final,colorlinks=true,linkcolor=blue,citecolor=blue,urlcolor=blue} 
\bibpunct{(}{)}{;}{a}{}{,} 

\makeatletter
\renewcommand*\aa@pageof{, page \thepage{} of \pageref*{LastPage}}
\makeatother

\begin{document}

\title{Constraints on the acceleration region of type III radio bursts from decimetric radio spikes and faint X-ray bursts}

\titlerunning{Constraints on the acceleration region of type III radio bursts from radio spikes and X-ray bursts}


\author{Sophie Musset \inst{1} \and Eduard Kontar \inst{1} \and Lindsay Glesener \inst{2} \and Nicole Vilmer \inst{3} \and Abdallah Hamini \inst{3}}
\institute{SUPA, School of Physics and Astronomy, University of Glasgow, Glasgow G12 8QQ, UK
\and School of Physics and Astronomy, University of Minnesota, 116 Church St SE, Minneapolis, MN 55455, USA
\and LESIA, Observatoire de Paris, Université PSL, CNRS, Sorbonne Université, Université de Paris, 5 place Jules Janssen, 92195 Meudon, France}

\date{Received date / Accepted date}

\abstract 
{We study the release of energy during the gradual phase of a flare, characterized by faint bursts of non-thermal hard X-ray (HXR) emission associated with decimetric radio spikes and type III radio bursts starting at high frequencies and extending to the heliosphere.
}
{We characterize the site of electron acceleration in the corona and study the radial evolution of radio source sizes in the high corona.}
{Imaging and spectroscopy of the HXR emission with \textit{Fermi} and RHESSI provide a diagnostic of the accelerated electrons in the corona as well as a lower limit on the height of the acceleration region. Radio observations in the decimetric range with the ORFEES spectrograph provide radio diagnostics close to the acceleration region. Radio spectro-imaging with LOFAR in the meter range provide the evolution of the radio source sizes with their distance from the Sun, in the high corona.}
{Non-thermal HXR bursts and radio spikes are well correlated on short timescales. 
The spectral index of non-thermal HXR emitting electrons is -4 and their number is about $2\times 10^{33}$ electrons/s. The density of the acceleration region is constrained between $1-5 \times 10^9$ cm$^{-3}$. Electrons accelerated upward rapidly become unstable to Langmuir wave production, leading to high starting frequencies of the type III radio bursts, and the elongation of the radio beam at its source is between 0.5 and 11.4 Mm. The radio source sizes and their gradient observed with LOFAR are larger than the expected size and gradient of the size of the electron beam, assuming it  follows the expansion of the magnetic flux tubes.
}
{These observations support the idea that the fragmentation of the radio emission into spikes is linked to the fragmentation of the acceleration process itself. The combination of HXR and radio diagnostics in the corona provides strong constrains on the site of electron acceleration. Spectroscopic imaging of radio sources with LOFAR provide further evidence that the radio source sizes are strongly affected by scattering of radio-waves.
}

\keywords{Sun; Solar flares; particle acceleration}

\maketitle 

\section{Introduction} \label{sec:intro}

The Sun is a powerful particle accelerator, being the source of energetic particles that have been detected in various locations within the heliosphere, and which are responsible for electromagnetic radiation in different ranges.
One common signature of these energetic electrons in the corona and interplanetary medium are type III radio bursts, characterized by a fast negative frequency drift, frequently observed at various wavelengths, from decimetric to kilometric waves \citep[see][for a review]{reid_2014_review}.
Type III bursts are believed to be generated by unstable beams of energetic electrons generating Langmuir waves at the local plasma frequency, which can be converted into electromagnetic at the fundamental and first harmonic of the plasma frequency \citep[as first proposed by][]{ginzburg_etal_1958}. 
The observed frequency drift of the bursts is therefore due to the propagation of the electron beams away from the Sun, in decreasing ambient plasma density.
The size of the electron beam is supposedly linked to the size of the magnetic flux tube in which they propagate, and therefore expands as they propagate upwards in the high corona and interplanetary medium.

Among the sources of energetic electrons are solar flares, characterized by a impulsive energy release in the corona. 
During solar flares, some of the dissipated energy accelerates particles. 
Energetic electrons propagating in the corona and further down in the solar atmosphere will produce bremsstrahlung emission in hard X-rays (HXR). 
Since HXR emission is produced in dense plasma, it is generally thought to be produced in closed magnetic loops located below the site of energy release and particle acceleration. 
Type III radio bursts are commonly observed simultaneously with flare-associated HXR emission, which supports the idea that flares are one of the sources of energetic electrons in the heliosphere.
The link between type III radio bursts and HXR emissions has been studied for many years \citep[see for instance][for a review]{pick_vilmer_2008}. 
\cite{kane_1972} performed one of the first studies of the link between HXR bursts and type III radio bursts, on 18 events. 
They showed the good time correlation between these signatures and deduced that energetic electrons responsible for both X-ray and radio emissions were accelerated simultaneously at a single acceleration site during the flare. 
These conclusions were confirmed by other statistical studies \citep{kane_1981,hamilton_etal_1990,aschwanden_etal_1995,arzner_benz_2005,reid_vilmer_2017}. However, these studies showed that only a fraction of type III bursts were associated with HXR bursts \citep{kane_1981,hamilton_etal_1990} and when there is a correlation, only a fraction of events show no time delay (or a time delay < 1s) between the radio and HXR emission \citep{arzner_benz_2005,reid_vilmer_2017}. 
\cite{reid_vilmer_2017} analyzed 10 years of joint observation with \textit{Reuven Ramaty High Energy Solar Spectroscopic Imager} (RHESSI) and the \textit{Nan\c cay Radio Heliograph} (NRH) to carry out a statistical study of the relation between coronal type III radio bursts (emission above 100 MHz) and their X-ray counterpart. 
A weak correlation was found between the radio flux below 327 MHz and the X-ray flux in the 25-50 keV range. 
They also observed that only 54 \% of their coronal type III bursts had a counterpart in the interplanetary medium, and that interplanetary type III radio bursts were preferably related to beams with more energetic electrons above 25 keV.

In-situ measurements of radio-producing electron beams showed that a moderate number of electrons ($10^{33}$ electrons in the 10-100 keV range) was sufficient to produce strong type III burst emission \citep{lin_1974}.
This number can be compared to the typical number of energetic electrons produced during a flare ($ \approx 10^{36}$) and is consistent with the idea that only a very small fraction of the accelerated electrons (0.1-1\%) escape the solar corona and reach the interplanetary medium \citep{lin_and_hudson_1971,krucker_etal_2007}. 
However, this fraction could be significantly higher in the case of small flaring events: \cite{james_etal_2017} studied 6 electron events detected in-situ  associated with faint HXR bursts, for which the number of X-ray producing electrons was between $3.6 \times 10^{30}$ and $1.8 \times 10^{33}$ electrons$/$s; for these events, the fraction of escaping electrons over HXR-emitting electrons ranged between 5 and 150\%.

The starting frequency of type III radio bursts seems to be particularly linked to the HXR emission: 
\cite{kane_1981} showed that starting frequencies correlated with the intensity of the impulsive X-ray bursts. 
They also noted that the correlation between X-ray and type III radio bursts was increasing with increasing starting frequency of the radio bursts. 
Type III radio bursts with high starting frequencies, for example in the decimeter range, are emitted in higher density plasma, and therefore closer to the acceleration site of energetic electrons. 
This explains the higher correlation with the HXR emission, which is itself produced close to the site of energy release.
\cite{reid_etal_2011,reid_etal_2014} investigated how the starting frequencies of radio bursts were linked to the properties of the electron beam. Combining radio observations with X-ray diagnostics of the energetic electrons, they showed that the starting frequency of type III radio bursts is strongly linked to the spectral index of the electron distribution and the elongation of the beam at its sources, which can be interpreted as the size of the acceleration region. They showed that harder electron spectra would result in higher starting frequencies of type III radio bursts. This suggests that with harder electron spectra, the electron beam becomes unstable to Langmuir wave production after a shorter propagation distance. The statistical study by \cite{reid_etal_2014} found values of the acceleration region height between 25 and 183 Mm, with typical sizes from 2 to 13 Mm, confirming that the acceleration of energetic electrons is happening in extended region high in the corona. The electron spectral index of the events presented in this study were in the range 4-8. It should be noted that the extent of the acceleration region here is estimated assuming that it corresponds to the length over which the instabilities of the electron beam develops to produce the radio emission. A study of type III radio bursts observed with the VLA by \cite{chen_etal_2018} calculated an upper limit on the size of the acceleration region of 600 km$^2$, which might indicate that the acceleration region has in fact a smaller extent than the elongation of the electron beam needed for the plasma instabilities to develop and for the production of the radio coherent emission.


Assuming that the higher the frequency of radio emissions, the closer to the acceleration region it is emitted, it is perhaps not surprising that the highest association with HXR of any kind of solar coherent radio emission is observed with radio spikes \citep{gudel_etal_1991,benz_and_kane_1986,aschwanden_and_gudel_1992}. 
Radio spikes are short-lived, narrow-band emissions that are observed at high frequencies, between 0.3 and 8 GHz; they exist on timescales of the order of tens of milliseconds and have a bandwidth of a few percents of the center frequency \citep{benz_1986,paesold_etal_2001}. 
They are often observed in clusters in association with type III radio bursts, with frequencies slightly higher than the starting frequencies of the type III radio bursts \citep{benz_etal_1992}. 
While radio spikes are often closely related with HXR emission, there is often no distinct temporal correlation between the X-ray flux and the radio spikes  \citep{benz_1985,benz_etal_2002}.
\cite{paesold_etal_2001} showed that the spikes emission location is consistent with the backward extrapolation of type III radio burst trajectories. On the other hand, \cite{benz_etal_2002} and \cite{battaglia_benz_2009} reported that the position of decimetric spike sources were significantly displaced from the hard X-ray sources (by 20 to 400 arcsec in the studied events), located higher in the corona, potentially close to the acceleration site.
The close temporal relationship with HXR bursts and spatial and temporal association with type III radio bursts of spikes led to the conclusion that spikes were closely related to the acceleration region of energetic electrons, resulting from plasma emission in a dense and small region of the solar upper atmosphere \citep{krucker_etal_1995,krucker_etal_1997}.
\cite{benz_1985} suggested that the fragmentation of the radio emission into spikes was due to a discontinuous exciter, which could be the signature of a discontinuous energy release (for instance, in a fragmented current sheet). Using the observed bandwidth of spikes, their source dimension was estimated to the order of 200 km, and interpreted as the typical dimension of numerous acceleration sites where current instabilities develop. 
Similarly, \cite{karlicky_barta_2011} showed that decimetric radio spikes originate from the fragmentation of the current sheet between merging plasmoids using a MHD simulation of the process. 

Only a few studies analyzed the short timescale ($< 2$ seconds) variations of the hard X-ray flux during solar flares, mainly because of instrumental limitations. 
Recently, sub-second variations of the X-ray flux were examined using high cadence observations from the spectrometer \textit{KONUS} on board of the \textit{Wind} spacecraft, and the \textit{Gamma-ray Burst Monitor} (GBM) on board the \textit{Fermi} spacecraft. 
\cite{altyntsev_etal_2019} used these observations during a flare to constrain the acceleration timescale, which is found to be smaller than 50 ms. \cite{knuth_and_glesener_2020} showed that the \textit{Fermi}/GBM observations were particularly adapted to the study of sub-seconds X-ray spikes for two case studies, in which X-ray spikes mean durations were 0.49 and 0.38 seconds.

Recent advanced radio-telescopes such as the \textit{LOw Frequency ARray} (LOFAR), the \textit{Murchison Widefield Array} (MWA) and the \textit{Karl G. Jansky Very Large Array} (VLA) provide imaging spectroscopy of the radio sources with orders of magnitude better resolution than before, providing improved observations of type III radio bursts and new constraints for theoretical models of the generation of electron beams, radio burst emission processes and radio-wave propagation effects \citep[see][for a review]{reid_2020}.
Radio-wave scattering on density fluctuations is one of the dominant process affecting the propagation of radio-waves in the heliosphere \citep{steinberg_etal_1984,steinberg_etal_1985,bastian_etal_1994,bastian_etal_1995}.
It has been shown to affect the observed sizes of radio sources and shift their apparent positions \citep[e.g.][]{steinberg_etal_1971}. \cite{chrysaphi_etal_2018} carefully examined type II burst emission observed with LOFAR and shown that the radio sources observed shift could be explained as an effect of radio-wave scattering.
Recent observations of fine structures in type III radio bursts with LOFAR were used to derive typical sizes of the electron beam and compared to the observed radio sources sizes, demonstrating that radio wave scattering was a dominant factor in determining the size of the radio sources observed in that frequency range \citep{kontar_nature_2017,sharykin_etal_2018}. Similar results have been shown for radio sources around 100 MHz, observed by the MWA \citep{mohan_etal_2019}.
Radio-wave scattering also affects the radio source positions, and the radio burst decay times: \cite{kontar_etal_2019} recently showed that the radio burst properties could only be explained with an anisotropic scattering of the radio-waves.

In this paper, we analyze a small event of particle acceleration which happens during the gradual decay phase of a more intense solar flare. 
This event is caracterized by a faint bursts of non-thermal X-ray flux, closely associated in time with decimetric radio spikes and type III radio bursts starting at frequencies as high as 600 MHz and extending to the interplanetary medium. 
We use hard X-ray emission and decimetric coherent emission as signatures of the energetic electron population at the source of the energetic electron beams. 
Observations are presented in section \ref{sec:observations}, and analyzed in section \ref{sec:analysis}. The implication regarding the geometry of the event, the link between X-ray and radio emitting electrons, the size of the acceleration region and the sizes of the radio sources in the high corona are discussed in section \ref{sec:discussion}.


\section{Observations and methods}
\label{sec:observations}

\subsection{Overview of the event}

This study focuses on radio and X-ray emission from the M3.7 class flare $\text{SOL2017-09-09T10:50}$ that happened on September 09 2017, from 10:50 to 11:42 and peaking around 11:04 UT. This flare originated from the active region 12673 which was located on the west limb of the Sun. 
This event was partially observed in hard X-rays by the \textit{Fermi} \textit{Gamma-ray Burst Monitor} \citep[GBM,][]{gbmpaper} and the \textit{Reuven Ramaty High Energy Solar Spectroscopic Imager} \citep[RHESSI,][]{rhessi}. It was also observed by the \textit{LOw Frequency ARray} \citep[LOFAR,][]{lofar} in tied-array mode, by the Nan\c cay Decametric Array  \citep[NDA,][]{nda}, by the solar radio spectrograph ORFEES (\textit{Observations Radio pour FEDOME -Fédération des Données Météorologiques l’Espace- et l’Etude des Eruptions Solaires}), and by the \textit{Radio and Plasma Wave} \citep[WAVES,][]{waves} instrument on the \textit{Wind} spacecraft. 

\begin{figure}
\includegraphics[width=\linewidth]{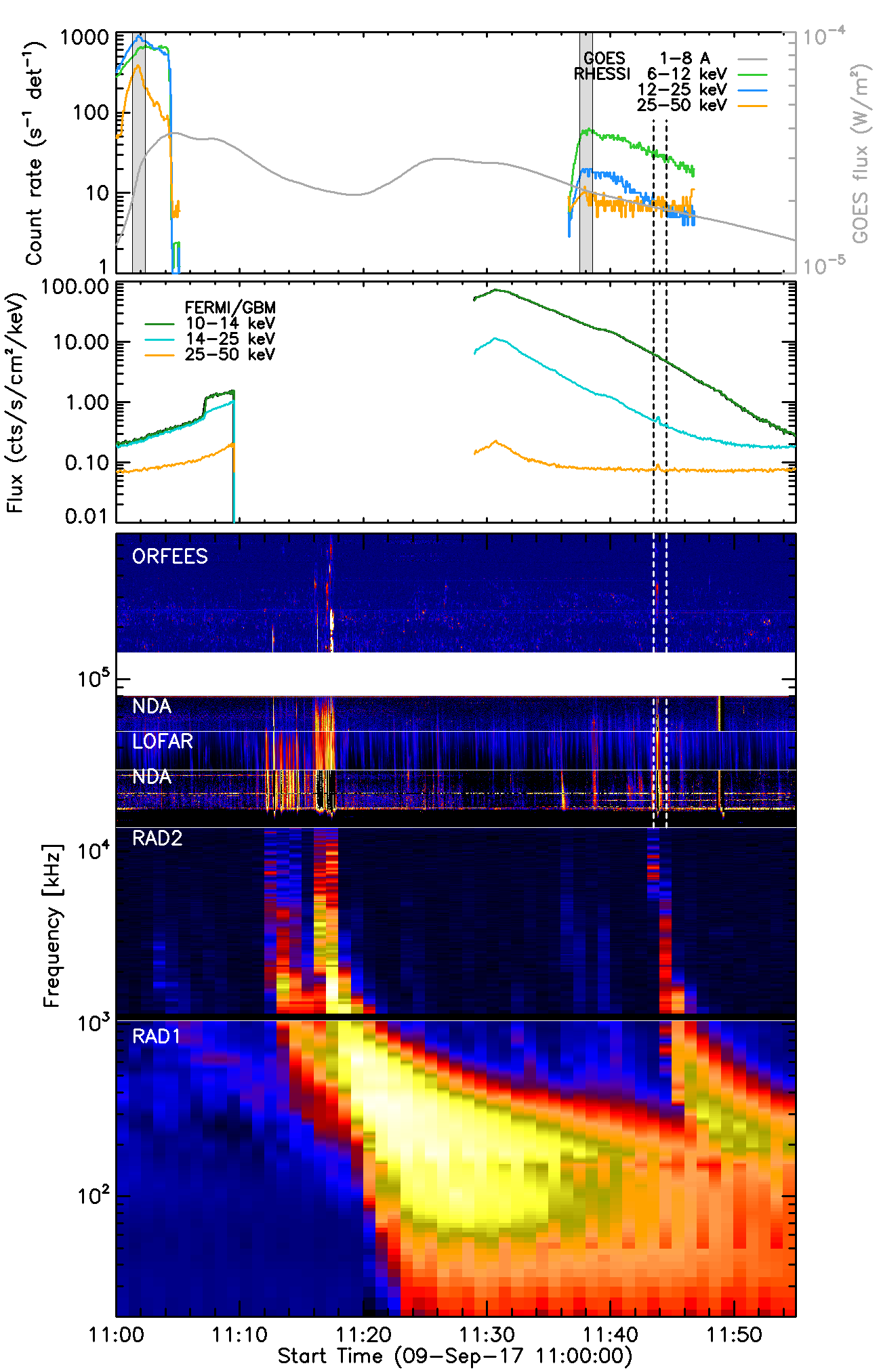}
\caption{Overview of the X-ray and radio emission analyzed in this paper. Top: RHESSI and GOES lightcurves, corrected from attenuator changes. Shaded intervals indicate intervals used for RHESSI imaging. Middle: \textit{Fermi}/GBM lightcurves. Bottom: Radio dynamic spectra, combining ORFEES (Nan\c cay, France), the \textit{Nan\c cay Decametric Array} (NDA), LOFAR and \textit{Wind}/WAVES radio observations (labeled RAD1 and RAD2). Vertical dashed lines indicate the time interval used for figure \ref{zoom}.}
\label{overview}
\end{figure}

\begin{figure}
\includegraphics[width=\linewidth]{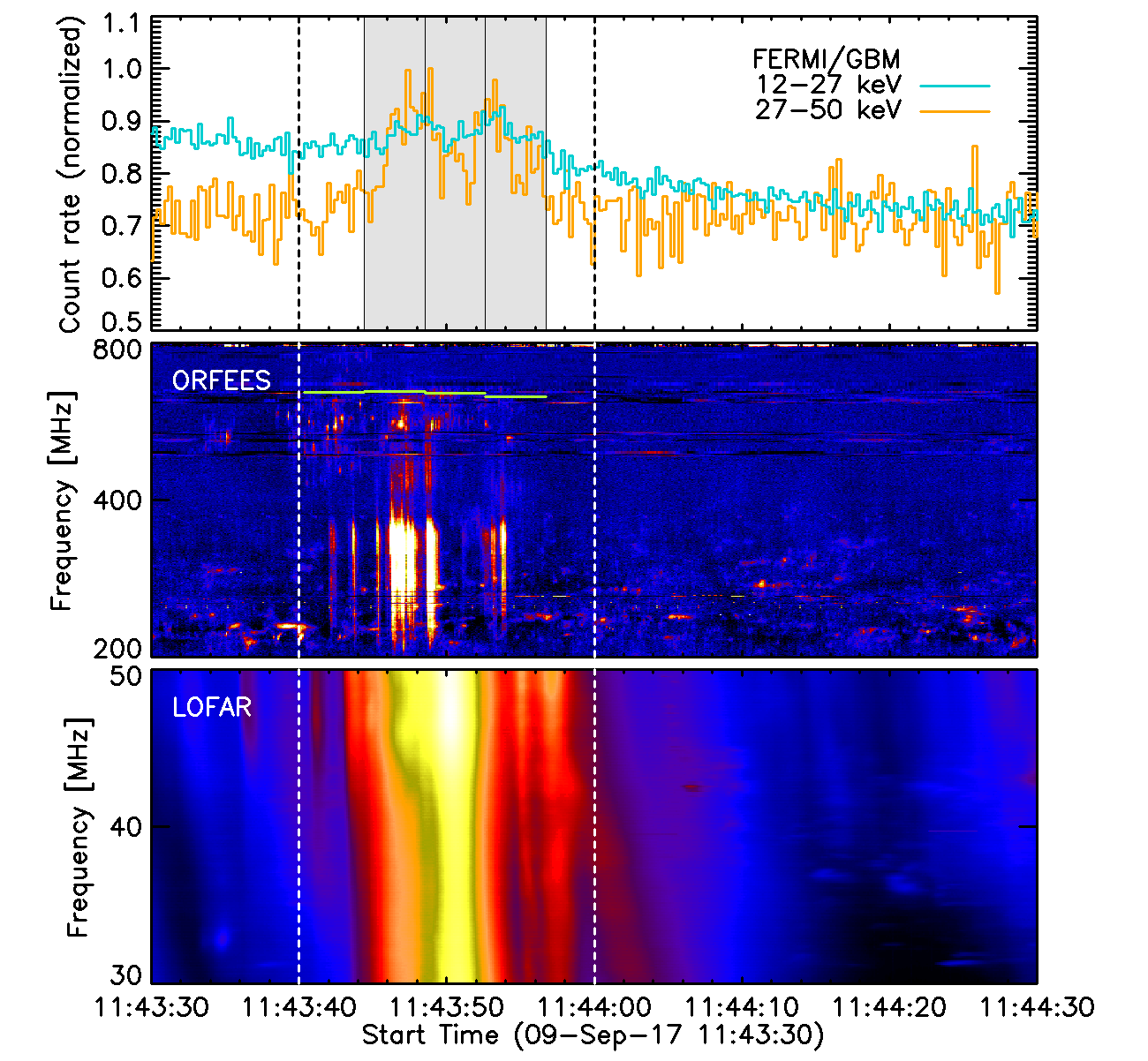}
\caption{60-second interval zoom on the enhanced radio and X-ray emission during the gradual phase of the flare. Top: normalized \textit{Fermi}/GBM lightcurves. Shaded intervals indicate the time intervals on which a non-thermal spectral component is detected. Middle: ORFEES dynamic spectrum. Green horizontal lines indicate the lower limit for the plasma frequency at the X-ray source as estimated from the X-ray spectral analysis and imaging (note that the X-ray images are produced 6 minutes before the spectral analysis and are used as a proxy for the emission at the considered time). Bottom: LOFAR dynamic spectrum. Vertical dashed lines are plotted to guide the eye on the start and end of the enhanced emission.}
\label{zoom}
\end{figure}

An overview of the whole flare seen by these instruments is displayed in figure \ref{overview}. 
In this figure, two main groups of fast, drifting radio emissions, characteristic of type III radio bursts, can be seen. These type III bursts extend to low frequencies (tens of kHz) as seen in the \textit{Wind}/WAVES observations, indicating that the energetic electrons escaped the solar atmosphere and propagated in the interplanetary medium.
The first group of type III bursts happens around 11:17 UT. At that time, \textit{Fermi} and RHESSI are both in night intervals. 
A second group of fainter type III radio bursts happens during the gradual phase of the flare around 11:44 UT. 
This group of type III emission happens at the time of a small increase in the X-ray flux between 15 and 50 keV, as seen in the \textit{Fermi}/GBM lightcurves. 
This paper focuses on this second group of type III radio bursts since it is particularly well correlated in time with the X-ray signatures, and these X-ray bursts are particularly faint, compared to earlier study cases used to link radio and X-ray signatures of electrons during solar flares. 
Observations for this group of type III bursts and the associated X-ray emission is shown on a one-minute time interval in figure \ref{zoom}.
This event provides the opportunity to characterize the source of the electron beams at the origin of the radio emission, since the absence of significant delays between the different signatures of energetic electrons facilitate its interpretation.

In the following paragraphs, we describe the X-ray spectral analysis and imaging, followed by the analysis of the ORFEES and LOFAR observations of the type III radio burst emission.

\subsection{X-ray data analysis}

While \textit{Fermi} is designed for the observation of faint, distant astronomical sources, it also detects solar emission.
Due to its low-Earth circular orbit, observation of the Sun is interrupted by night periods every 90 minutes. 
The GBM has 12 detectors placed in order to cover the whole sky, sensitive to X-ray and gamma-ray radiation between 8 keV and 40 MeV. It provides continuous low spectral resolution data at a nominal cadence of 0.256 second and higher spectral resolution data with a nominal cadence of 4.096 seconds \citep{gbmpaper}.

For this event, the peak of the flare was missed by \textit{Fermi}, which was behind the Earth from 11:09:34 to 11:28:57 UT. In the later phase of the flare, the data from the four most sunward GBM detectors were combined to produce the GBM lightcurves displayed in figures \ref{overview} and \ref{zoom}. In the latter, using the high cadence measurements, a small increase in intensity in the 14-27 and 27-50 keV channels is noticeable at the time of the radio type III burst emission: two main peaks in intensity can be distinguished between 11:43:40 and 11:44:00 UT (between the two vertical dashed lines in figure \ref{zoom}). 
These high-energy X-ray peaks are very faint compared to the background X-ray emission of the gradual phase of the flare.

\subsubsection{Non-thermal excess emission around 11:43:48 UT}

\begin{figure}
\includegraphics[width=\linewidth]{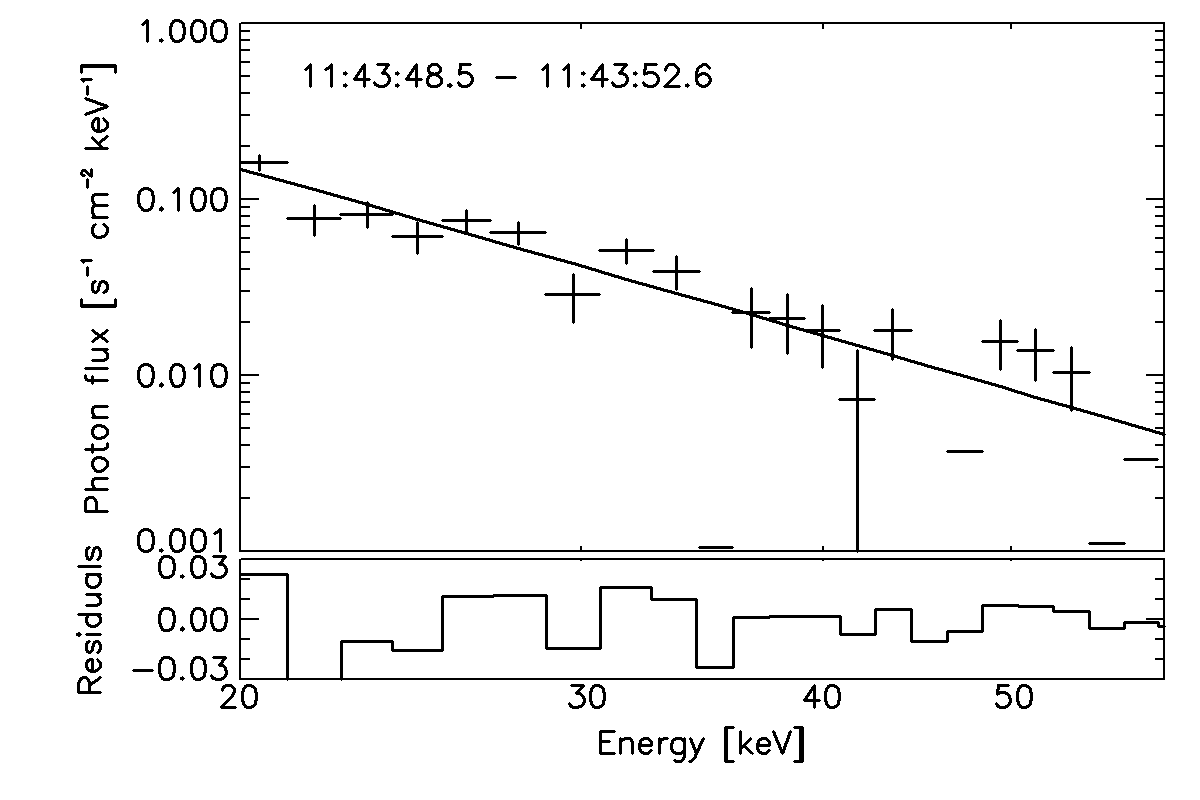}
\caption{Top: \textit{Fermi} photon flux spectrum between 11:43:49 to 11:43:53 UT. The background emission from the gradual phase of the flare has been removed. Data points are shown as crosses. The result of the fit, a single power law thick target model, is shown as a line. Bottom: Residuals of the spectral fit.}
\label{fermispectre}
\end{figure}

A spectral analysis of the GBM X-ray emission has been performed for this X-ray burst, between 11:43:44 and 11:43:56 UT, in three time intervals shown in gray in the top panel of figure \ref{zoom}. The spectral analysis is performed using the high spectral resolution data from \textit{Fermi}/GBM, and therefore at the nominal cadence of 4.096 seconds (not the high-time cadence of the lightcurves shown in figure \ref{zoom}).
The X-ray flux from the most sunward detector at the time of interest, Na\_01, is used. In order to characterize this particular excess of X-ray emission of X-ray emission (in the 15-50 keV range) above the decreasing X-ray emission of the gradual phase of the flare, we subtracted the flare emission, estimated just before 11:43:36 and after 11:44:09 UT, and linearly interpolated between these times. The spectrum of the remaining flux was fitted with a single power law thick-target model in OSPEX. The best fit result was obtained for the second time interval, and is presented in figure \ref{fermispectre}. The result of the fit to the spectrum gives an spectral index of 4.0 $\pm$ 0.3 for the electron spectrum, and an electron flux of $(2.0 \pm 2.3) \times 10^{33}$ electrons/s.

\subsubsection{Thermal X-ray spectral analysis}

We additionally performed the spectral analysis of the X-ray emission of the flare from \textit{Fermi}/GBM between 11:43:36 and 11:44:01 UT (divided into six time intervals). 
We first fitted a thermal model to the spectra. 
For the three time intervals between 11:43:44 and 11:43:56, a thermal component was not sufficient to correctly fit the data, and we added a thick target component as described in the previous paragraph with parameters fixed to the values determined by this previous analysis. 
This thermal component of the spectral fit between 11:43:36 and 11:44:01 UT results with emission measure values from 4.6 to 5.3 $\times 10^{47}$ cm$^{-3}$, while the plasma temperature stays around 17 MK. 

\subsubsection{X-ray imaging}

RHESSI was a solar spectro-imager sensitive in the 3 keV - 17 MeV range, operating between 2002 and 2018.
It was equipped with nine detectors are photon-counting detectors associated with collimators modulating the signal as the spacecraft rotated with a 4-second period. 
This modulation is used to reconstruct images of the emission sources \citep{rhessi_imaging}.
Similarly to \textit{Fermi}, RHESSI was in a low-Earth orbit, with periods of nights. 
Observations were also interrupted when the spacecraft passed through the South Atlantic Anomaly. 

\begin{figure}
\includegraphics[width=\linewidth]{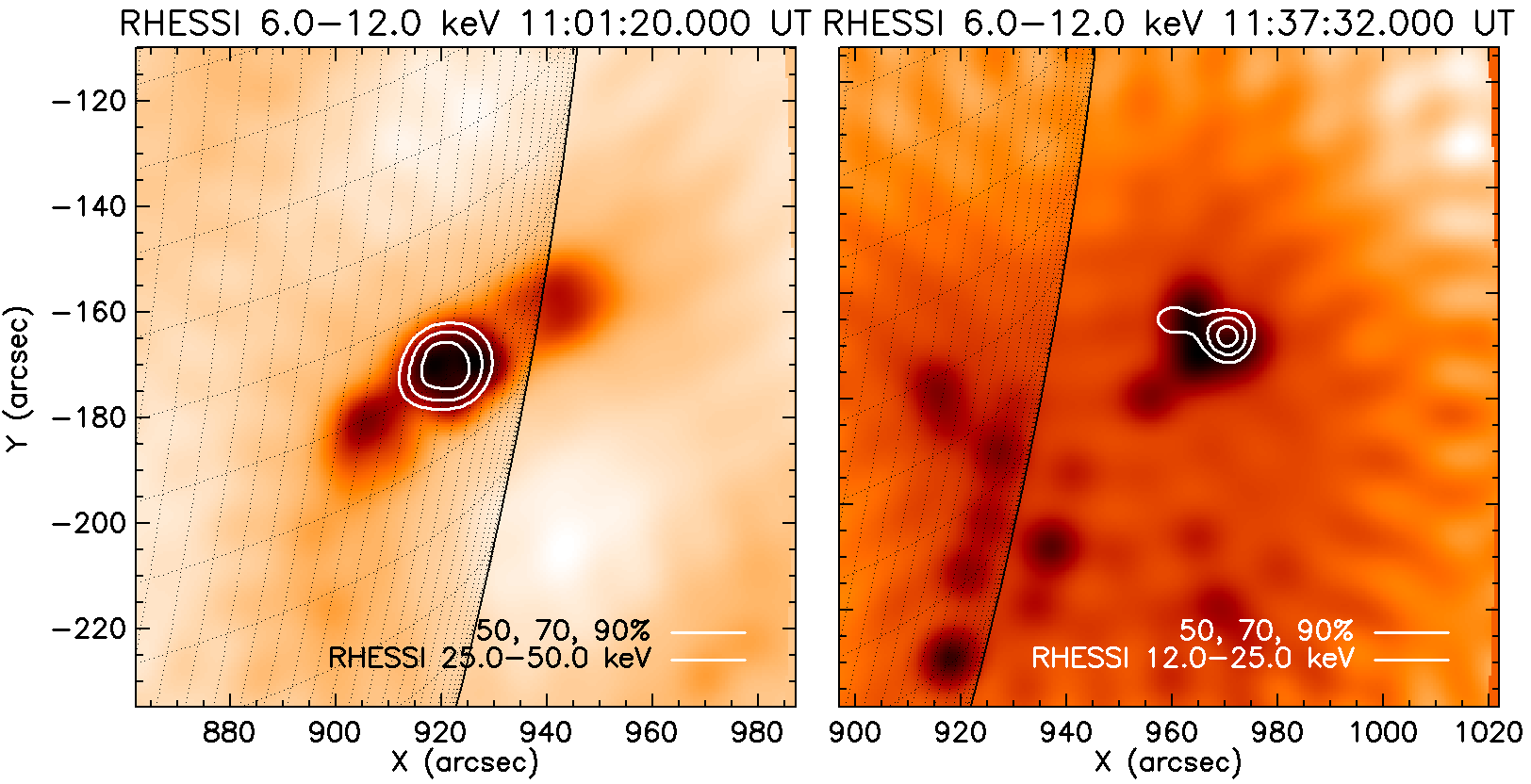}
\caption{RHESSI images at the beginning and end of the flare, reconstructed with the CLEAN algorithm using detectors 3, 6 and 8. Images at 6-12 keV are shown. The contours of the higher energy band emission (25-50 keV on the left image, 12-25 keV on the right) are shown in white.}
\label{rhessi_images}
\end{figure}

RHESSI was in an eclipse interval from 11:05:12 to 11:36:36 UT and therefore missed the peak of the flare. 
The satellite passed over the South America Anomaly (SAA) and stopped observing after 11:46:48 UT.
The RHESSI lightcurves are shown in figure \ref{overview}. 
These lightcurves are corrected for attenuator changes that happened at 11:00:44, 11:01:12, 11:37:00, and 11:37:28 UT with the routine provided in the SolarSoft.
Imaging of the X-ray emission was performed before and after the eclipse interval, providing the location and evolution of the X-ray source during the flare. The intervals corresponding to the images presented in figure \ref{rhessi_images} are shown in gray in the top panel of figure \ref{overview}.

This flare happened late during the RHESSI mission, and only a subset of detectors was operational at the time of the flare: detectors 1, 3, 6 and 8. 
The images were reconstructed with detectors 3, 6 and 8 since including detector 1 (sensitive to the smallest spatial scales but with a low signal to noise ratio) was only adding noise to the image. 
We used the CLEAN algorithm with a beam factor of 1.5 to produce the images that were integrated over 60 second time intervals. 
For the first image in figure \ref{rhessi_images}, produced at the beginning of the flare and before the RHESSI night, at 11:01:20 UT, it was possible to image the X-ray emission in the 6-12, 12-25 and 25-50 keV energy ranges. 
For the second image, after the RHESSI night, at 11:37:32 UT, the X-ray emission is fainter and it was possible to image only the 6-12 and 12-25 keV emission. 
Both images are shown in figure \ref{rhessi_images}. 
We do not show images at the time of non-thermal X-ray emission excess (11:43:30 to 11:44:30) because these images are particularly noisy due to poor photon statistics. However, the CLEAN algorithm during the 11:43:30 to 11:44:30 resulted with a few CLEAN components locations consistent with the locations of the X-ray emission shown in the second image in figure \ref{rhessi_images}. The location of the X-ray source shown in this image is therefore still present at the time of the faint non-thermal X-ray bursts, with a variation of its location lower than 10 arcsec.
The size of the X-ray source is defined at the area within the 50 \% intensity level. 
For the late emission (second image), the 6-12 keV source size is 0.12 arcmin$^2$. We use this measurement rather than a measurement on a later image, because the poor quality of the subsequent images increases the uncertainty over the size of the sources.

\subsection{ORFEES data analysis}

ORFEES is a radio spectrograph dedicated to solar observations at the radio station of Nan\c cay, in France. 
It provides spectral observations between 144 and 1000 MHz with a time resolution of 0.1 seconds. For this event, we used observations in the 200-800 MHz range.
We used to raw, full-cadence dynamic spectrum and subtracted the background using the routines provided by the instrument team. 
As can be seen in the middle panel of figure \ref{zoom}, between 11:43:40 and 11:44:00 UT, fast drifting features are visible in the 200-500 MHz range, interpreted as type III radio bursts. 

\begin{figure*}
\includegraphics[width=\linewidth]{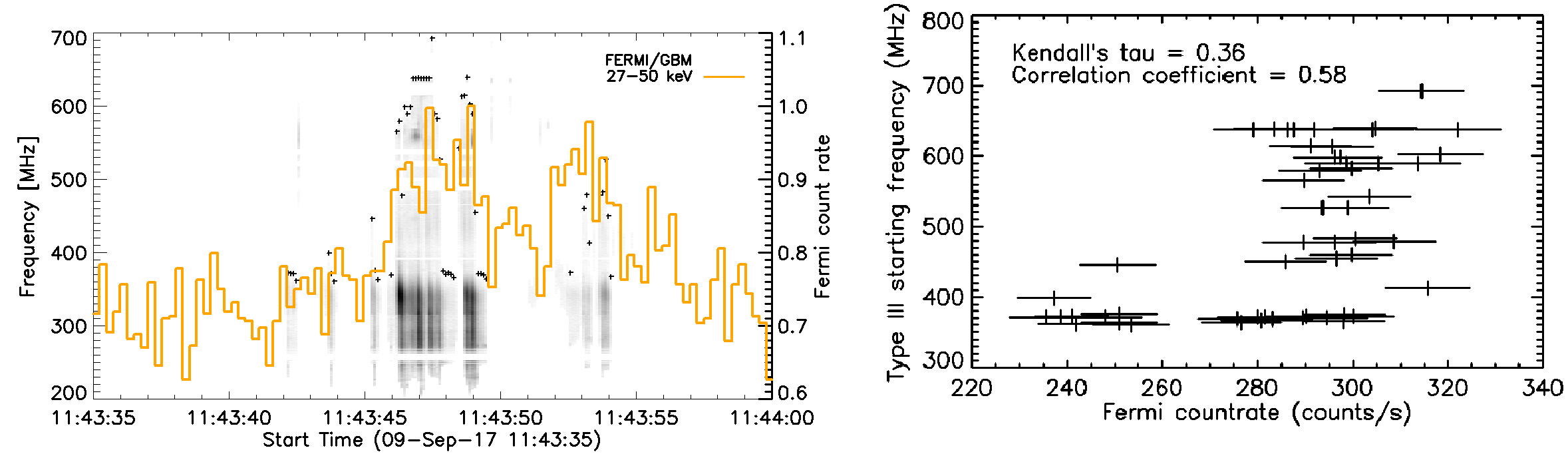}
\caption{Left: radio intensity for type III radio burst emission detected in the ORFEES spectrogram, as described in the text (gray scale data), and normalized X-ray flux at 27-50 keV seen by \textit{Fermi} (orange line). Starting frequencies of the type III radio bursts are showns at $+$ symbols in the plot. Right: correlation between the X-ray count flux and the starting frequency of the radio bursts.}
\label{typeIII_fermi}
\end{figure*}

After removing bad channels in the spectrogram, we calculated the background in each frequency channel during quiet time intervals and identified the radio burst emission above the background. The starting frequencies of the individual bursts evolve between 355 and 693 MHz from 11:43:42 to 11:43:54.
The detected bursts are shown over the \textit{Fermi} X-ray lightcurve in the left panel of figure \ref{typeIII_fermi}. It can be seen that the occurrences of type III radio bursts clearly correspond to the excess X-ray emission seen by \textit{Fermi}. 
One can also clearly see a low-high-low trend of the starting frequency of type III radio bursts during the event. This trend has been observed in many other type III bursts events related to X-ray flare emission of larger flux \citep{reid_etal_2014}. 
Furthermore, there is a weak correlation between the X-ray count rate and the starting frequency of the type III radio bursts, as shown in the right panel of figure \ref{typeIII_fermi}. The low time cadence available for spectroscopy unfortunately prevents the analysis of the correlation between the X-ray spectral index and the starting frequencies.
To characterize this correlation, we used the nonparametric Kendall's tau coefficient, based on the comparison of ranks in the two set of values. 
This method tests the correlation between two sets of values without assumption on the form of the correlation, and in our case the coefficient is of 0.36, indicative of a weak correlation. 
Assuming that we have a linear correlation between the type III radio bursts starting frequency and the X-ray count rate, the correlation coefficient is 0.58.

\begin{figure}
\includegraphics[width=\linewidth]{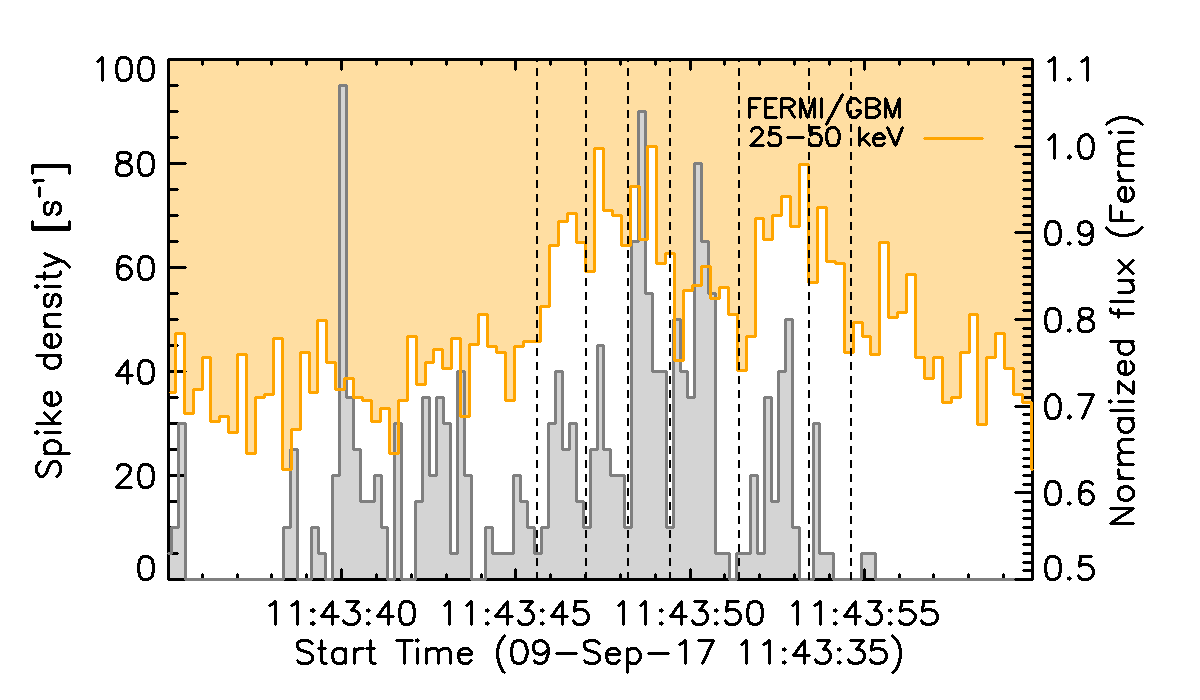}
\caption{Left: Density of radio spikes between 538-640 MHz per second (black), compared to the normalized X-ray flux at 27-50 keV observed by \textit{Fermi} (orange). Vertical dashed lines point to local minima in the X-ray lightcurves that correspond in time with local minima in the density of radio spikes.}
\label{spikes_fermi}
\end{figure}

Around the same time as the group of type III radio bursts described above, between 538 and 640 MHz, some short-lived and narrow features are visible, identified as to radio spikes. 
We identify individual radio spikes similarly to the method used to detect the type III radio burst: in the restricted region of the dynamic spectrum where spikes are seen, we automatically detect groups of adjacent pixels with intensity above the background, reject the large groups associated with the radio bursts, and keep the remaining small groups of pixels as the radio spikes.
These detected spikes have an average frequency width of 3.3 MHz (0.6\% of the center frequency on average), and exist on sub-second time scales of an average duration of 0.14 sec. Note that the ORFEES data has a 0.1 time cadence, and that some of the spikes might not be resolved in time. 
The density of spikes evolution in time is compared to the non-thermal X-ray lightcurves in figure \ref{spikes_fermi}. 
As can be seen on this figure, spikes start to be observed a few seconds before the detection of a rise in the X-ray lightcurve in the 27-50 keV range, but correlate in general well with the X-ray lightcurve. In particular, the spike density and the X-ray flux share similar fine scale structures: local minima in the density of spike seem to correlate well with local minima of the X-ray flux, as marked by the dashed vertical lines in figure \ref{spikes_fermi}.

\subsection{LOFAR data analysis}
\label{ssec:LOFARanalysis}

The LOFAR radio-telescope is an astronomical telescope used for solar observations during dedicated campaigns. 
It is composed of several arrays in stations located in Europe, with the core stations located in the Netherlands. 
While this instrument is an interferometer, producing interferometric visibilities limits the imaging cadence. 
Instead, data from the LOFAR core stations can be combined into multiple tied-array beams, which offer the possibility to image the Sun with the time cadence of the dynamic spectra \citep{lofar,kontar_nature_2017}.

For this observation, LOFAR provided data in the 30-50 MHz range with a 0.67 second time cadence and a frequency resolution of 200 kHz, using 216 beams. The LOFAR dynamic spectra is seen in figures \ref{overview} and \ref{zoom}.
In tied-array mode, the mosaic of beams is pointed at the Sun in a honeycomb pattern. Images can be reconstructed by interpolating in the plane of the sky the intensity measured in each beam. 
The reconstructed images are the convolution of radio sources with the telescope beam, which has an area of 142 arcmin$^2$ at 30 MHz.

We use the 50\% intensity contours of these images to calculate the radio source position, size and area, similar to the observations described in \cite{kontar_etal_2019} and explained in appendix \ref{app:contours}. 
We also fitted the beam fluxes with an elliptical Gaussian model for the image source as described in \cite{kontar_nature_2017} and presented in appendix \ref{app:fit}, and use the result of the fit as a measurement of the source position, size and area. 
We found a general agreement between those two methods to derive the radio source size properties. In the following we use the results derived from the elliptical Gaussian model for the sources sizes and positions.

\begin{figure}
\includegraphics[width=\linewidth]{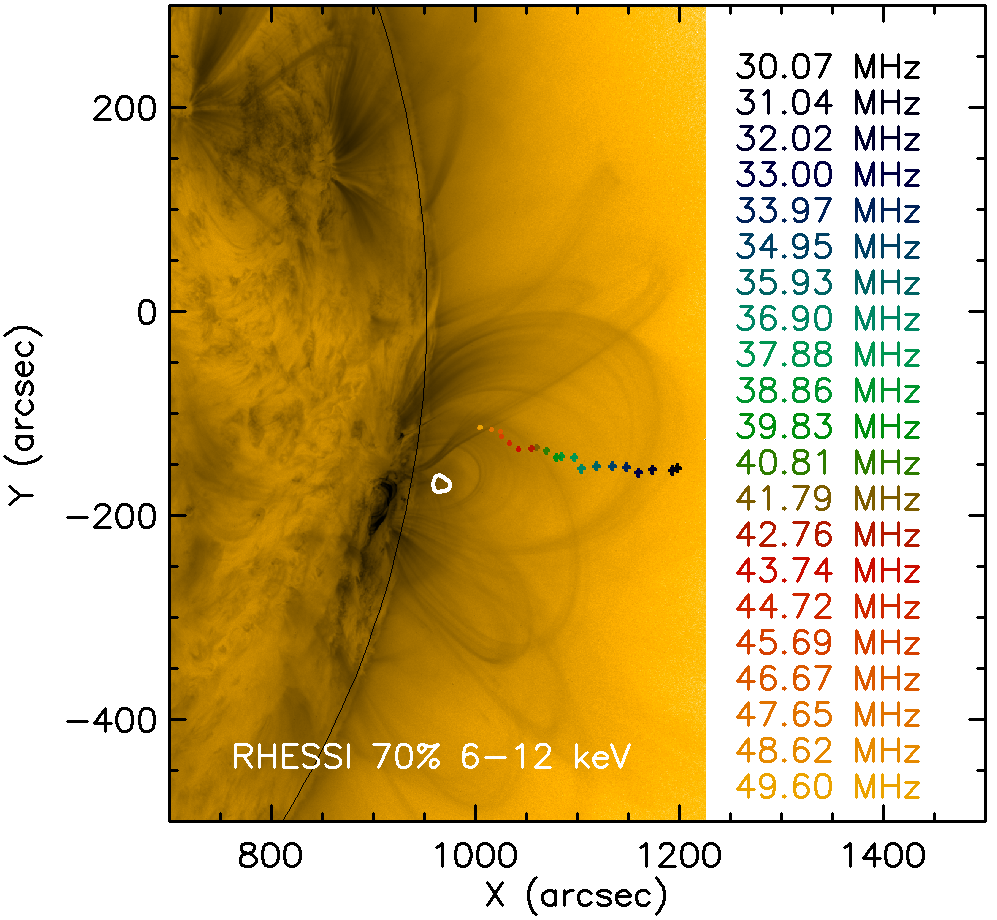}
\caption{Radio source centroid positions calculated at 11:43:50 UT overplotted on AIA image at 171 A at 11:43:45 UT. White contours show the 70\% contour of RHESSI emission at 6-12 keV at 11:38:02 UT.}
\label{centroidpos}
\end{figure}

The centroid positions deduced from the elliptical Gaussian fit are shown in figure \ref{centroidpos}, superimposed on an EUV image of the active region. 
The radio source evolves outwards from the active region with decreasing frequency, as expected for coherent plasma emission.

\begin{figure}
\includegraphics[width=\linewidth]{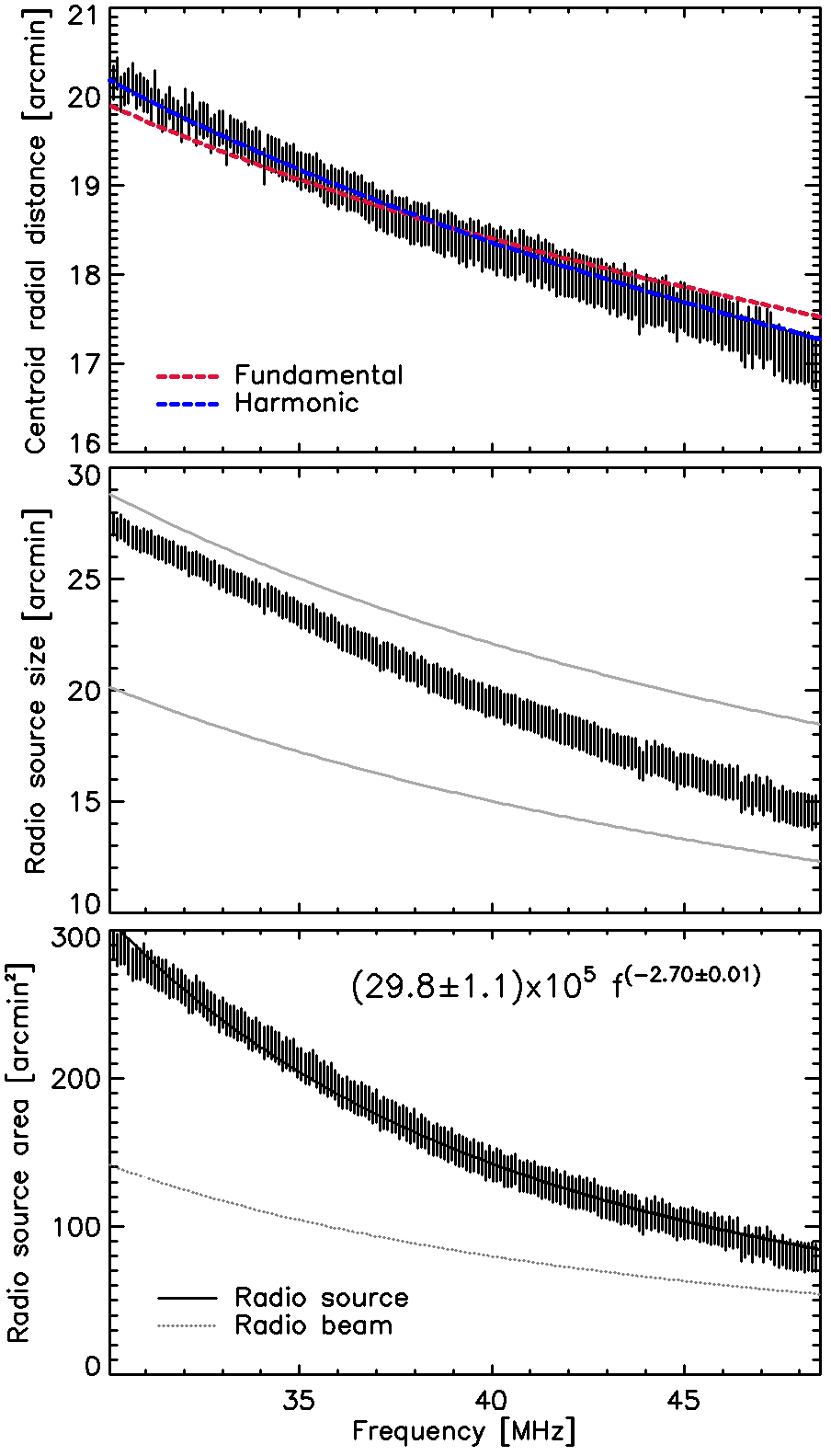}
\caption{Evolution of radio source parameters with frequency. Top: Source centroid radial position in arcminutes as a function of frequency. Overplotted are predictions of the source position assuming that the radio is emitted at the fundamental or harmonic frequencies, in red or blue respectively (see text for details). Middle: Source size in arcminutes as a function of frequency. The shaded area shows the expected evolution of the source size derived from a fit to different observations compiled by \cite{kontar_etal_2019}. Bottom: Source area in arcmin$^2$ as a function of frequency. The beam area (shown bu the dotted line) has been subtracted from the observed area. A power-law is fitted to the distribution and the result of the fit is shown a thick black line. }
\label{sourcevsfreq}
\end{figure}

The evolution of the radio source radial distance, size (FWHM) and area, deduced from the elliptical Gaussian fit to the radio flux, is shown in figure \ref{sourcevsfreq}. In the top panel, the centroid radial distance of the source is shown decreasing with increasing frequency: this is the radial distance of the radio source projected in the plane of sky. 
Using the density model described in section \ref{sec:densitymodel}, we can evaluate the plasma density as a function of the radio source radial distance from the Sun. 
The plasma frequency $f_p$ is directly related to the plasma density $n_e$ with the approximate relation:
\begin{equation}
f_p = 9 \sqrt{n_e}
\label{eqn:fp_ne}
\end{equation}
where $n_e$ is in cm$^{-3}$ and $f_p$ is in kHz. Radio coherent emission is generally emitted at the plasma frequency (fundamental emission) or twice the plasma frequency (harmonic emission).
Assuming that the electron beams propagate radially from the Sun in a direction that has an angle $\phi$ with the plane of sky, the true radial distance of the radio source $d_S$ relates to the observed centroid distance $d_C = d_S \cos \phi$. 
The observed evolution of the radial distance is therefore controlled by the density gradient, while $\cos \phi$ is a normalization factor.
With these assumptions, we overplot on the data in the top panel of figure \ref{sourcevsfreq} the evolution of frequency of the projected distance of a radio source emitted at the fundamental (harmonic) frequency, in a direction characterized by $\phi = 45^{\circ}$ ($\phi = 55^{\circ}$). Both curves are in agreement with the observations; the assumption that the radio emission is harmonic emission provides an assumption slightly closer to the observations than the assumption of fundamental emission.

The size of the radio source, here characterized by the FWHM of the elliptical Gaussian fit in major axis direction, is also decreasing with frequency, as shown in the middle panel of figure \ref{sourcevsfreq}. This evolution is consistent with typical radio source size evolutions as described in \cite{kontar_etal_2019}, where several studies of radio source sizes in different frequency ranges were combined, and the source size $\sigma_S$ in degrees as a function of the frequency $f$ in MHz was fitted with a power law distribution, providing the following trend:
\begin{equation}
\sigma_S = (11.78 \pm 0.06) \times f^{-0.98 \pm 0.05}
\end{equation}
This relation is shown, within its uncertainties, by the gray area in the panel.

Finally, the area of the radio source is shown as a function of frequency in the bottom panel of figure \ref{sourcevsfreq}. The radio beam area is also shown and is negligible compared to the size of the source. The observations were fitted with a power-law distribution, leading to a description of the area of the source $A_S$ in arcmin$^2$ as a function of frequency $f$ in MHz:
\begin{equation}
A_S = (19.3 \pm 0.5) \times 10^5 \times f^{-2.46 \pm 0.01}
\end{equation}

\section{Analysis}
\label{sec:analysis}

\subsection{Plasma density and height of the X-ray source}

Combining information from X-ray spectral fitting and imaging provides the opportunity to calculate the plasma density in the X-ray source.
Unfortunately, for the present event, photon statistics at the time of the bursts in the RHESSI data is to low to produce reliable images.
In the following, we use the closest X-ray image in time with acceptable quality to produce an estimate of the source size. 
We note that the thermal X-ray emission stays at the same location between 11:37:32 and 11:44:32, with decreasing intensity over time. 
Since the X-ray source in the second X-ray image (see figure \ref{rhessi_images}) is seen over the limb, we can make an approximation of the source height in the corona, assuming that projection effects can be neglected when the source is close to the solar limb. This value is a lower limit of the height of the source. In this image, the center of the source appears to be 19 Mm above the solar surface.

The size of the X-ray source is approximated by the surface of the image within the contour of 50 \% of the maximum. During the second time interval used for imaging, the 6-12 keV source cover a surface of 0.12 acrmin$^2$. Using this area $A$ and the emission measure $EM$ deduced from the X-ray spectral analysis, one can estimate the mean density $\overline{n}$ of the ambient plasma within the X-ray source:
\begin{equation}
\overline{n} = \sqrt{\frac{EM}{A^{3/2}}}
\end{equation}
This leads to a mean density of $8 \times 10^9$ cm$^{-3}$.
This estimation of the plasma density is subject to the uncertainty on the emission measure provided by the X-ray spectral analysis and the uncertainty on the source size. Assuming that we have a factor 2 of uncertainty on the source size, we can therefore derive upper and lower limits on this density: $5 \times 10^9$ to $13 \times 10^9$ cm$^{-3}$.

Using the lower limit of the plasma density and equation \ref{eqn:fp_ne}, the lower limit of the plasma frequency in the X-ray source is calculated for the time intervals in which radio spikes are visible in the ORFEES spectrograms, and is displayed as green horizontal lines in the middle panel of figure \ref{zoom}. The average value is 638 MHz. As can be seen in the figure, it is slightly higher than, or on the edge of, the frequency range in which radio spikes are visible.

\subsection{Heights of radio and X-ray sources: density models}
\label{sec:densitymodel}

By modeling the density radial evolution in the solar corona, it is possible to relate the height of the X-ray source to the plasma frequency, using the relation between the plasma frequency and the plasma electron density described by equation \ref{eqn:fp_ne}.
Historic density models such as the Newkirk \citep{newkirk_1967}, Baumbach-Allen \citep{allen_1947} and Saito \citep{saito_etal_1970} models fail to account for the high plasma density in flaring structures in the inner corona, the density gradient being to low. 
To interpret our observations, we need a density model that can  account for the starting frequency of the radio emission which is observed as high as 693 MHz, which corresponds to a density of $5.9 \times 10^9$ cm$^{-3}$ if the radio emission is emitted at the plasma frequency, or $1.5 \times 10^9$ cm$^{-3}$ in the case of harmonic emission. 
\cite{reid_etal_2014} addressed this issue using an exponential model in which the density gradient is controlled by a typical length scale. 
They calculated the value of this parameter to fit the observations of type III radio bursts presented in \cite{saint-hilaire_etal_2013}. 
However, this model fails to reproduce the density gradients further away from the Sun, where the density gradients are significantly different than the gradients in the low corona, and in particular here fails to explain the observed evolution of the radio source radial distance with frequency observed with LOFAR and presented in the top panel of figure \ref{sourcevsfreq}.

Therefore, we used a model described in \cite{kontar_etal_2019}:
\begin{equation}
n(r) = 4.8 \times 10^9 \left(\frac{R_s}{r}\right)^{14} + 3 \times 10^8 \left(\frac{R_s}{r}\right)^{6} + 1.4 \times 10^6 \left(\frac{R_s}{r}\right)^{2.3}
\label{eqn:densitymodel}
\end{equation}
were $n(r)$ is in cm$^{-3}$ and $R_s$ is the solar radius.
This model reproduces well the evolution of radial distances of the radio sources observed by LOFAR, as shown in the top panel of figure \ref{sourcevsfreq}. It also has a steep density gradient closer to the Sun enabling to predict high plasma frequencies at the flare site. We used this model to interpret the radio frequencies as heights in the corona, as shown in figure \ref{vsheight} and discussed in the rest of the paper.

\subsection{Size of the electron cloud generating type III radio bursts}

As discussed in \cite{reid_etal_2011}, the starting frequency of the radio type III bursts and the properties of the energetic electrons at their source are linked with the following relation: 
\begin{equation}
h_{typeIII} = \alpha d + h_{acc}
\end{equation}
where $h_{typeIII}$ is the height at which the type III emission starts, $h_{acc}$ is the height of the acceleration region, $d$ is the elongation of the energetic electron cloud at the origin of the beam of electrons producing the type III emission, and $\alpha$ is the electron spectral index of the velocity distribution of energetic electrons. 

From our observations, we can deduce values for $h_{typeIII}$, $h_{acc}$ and $\alpha$ and use them to estimate the characteristic size $d$.
The electron spectral index found from the X-ray spectral analysis is 4, which gives $\alpha = 8$ (in velocity space).

The starting frequency of radio type III bursts is around 620 MHz during the first peak of X-ray emission (11:43:44 to 11:43:50) and around 490 MHz during the second peak (11:43:53-11:43:54). Assuming harmonic emission, and using the density model represented by equation \ref{eqn:densitymodel}, these frequencies lead to densities of $1.2 \times 10^9$ and $0.7 \times 10^9$ cm$^{-3}$, and heights of 81 and 110 Mm, respectively.

The location of the thermal X-ray emission at the time of the faint X-ray bursts remains the same as the location of the source in the image reconstructed a few minutes earlier. 
This X-ray source is associated to the hot plasma that was heated during the main flare.
We consider that the height of the X-ray source provides a lower limit on the height of the acceleration region, assuming that the acceleration sites are located above these post-flare loops filled with hot plasma; 
and the height of the X-ray emission here is 19 Mm above the solar surface. 
Similarly, we consider that the electron density of the X-ray source, in the range $5-13 \times 10^9$ cm$^{-3}$, is the higher limit for the acceleration site electron density. 
We also assume that the radio spikes observed at high frequency in the ORFEES observations are closely related with the acceleration region: in this event, they are emitted between 538 and 640 MHz, very close to the starting frequency of type III radio bursts, and we adopt the emission at 640 MHz as an upper limit of the density (and thus lower limit on height) of the acceleration region. 
In the assumption that the radio emission from spikes is emitted at the plasma frequency, this emission corresponds to a density of $5.1 \times 10^9$ cm$^{-3}$ and a height of 0.5 Mm above the atmosphere using the density model represented by equation \ref{eqn:densitymodel}. 
This is below the height of the X-ray source, which contradicts our assumption that X-rays are produced below the acceleration region. 
This could show the limit of the validity of the density model that we are using, or it could be an indication that the radio emission here is emitted at the first harmonic. 
In that case, the density would be $1.3 \times 10^9$ cm$^{-3}$ and the height around 77 Mm above the solar surface.

Keeping a range of 19-77 Mm for the height of the acceleration region $h_{acc}$, we therefore calculate the range of the electron beam size $d$ to be 0.5-7.8 Mm during the first X-ray peak and 4.1-11.4 Mm during the second. 


\subsection{Evolution of radio source sizes between 30-50 MHz}

\begin{figure}
\includegraphics[width=\linewidth]{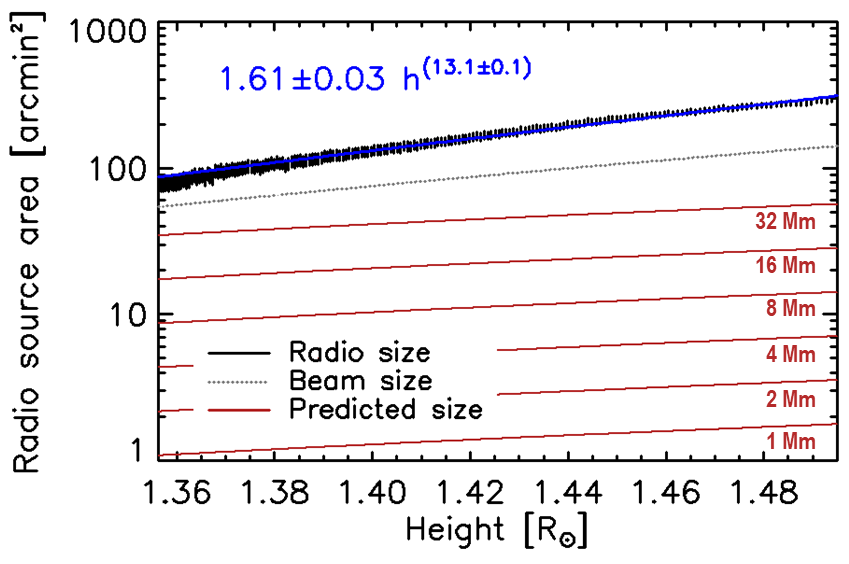}
\caption{Evolution of LOFAR source area as a function of radial distance (height) from the Sun (black). The radio beam area, shown as a dotted gray line, has been subtracted to the observed area. The blue curve shows the result of a power law model to the data. The red lines show the predicted size of the radio-producing electron beam, for different characteristic size of the radio-producing electron beam between 1 and 32 Mm, and for a source at a height of 19 Mm. The predicted size is based on a magnetic field expansion following the model of \cite{dulk_coronal_1978}.}
\label{vsheight}
\end{figure}

Using the observed evolution of the radio sources sizes as a function of frequencies (bottom panel of figure \ref{sourcevsfreq}) and the density model described in section \ref{sec:densitymodel}, we can study the evolution of the source size as a function of radial distance from the Sun. 
If we assume that the radio emission in this case is emitted at the first harmonic of the plasma density, we obtain the evolution shown in figure \ref{vsheight}.
If the size of radio sources is controlled by the size of the electron beam at the origin of the radio emission, it can be assumed that the evolution of radio source size with height follows the expansion of the magnetic flux tube containing the energetic electron beam, the magnetic flux being conserved. 
This assumption is supported by observations of electron beams emitting on flux tubes, as shown for example by \cite{klein_etal_2008}. 
An empirical approximation of the evolution of the magnetic field with height in the corona has been derived by \cite{dulk_coronal_1978}:
\begin{equation}
B = 0.5 \left( \frac{R}{R_s}-1 \right)^{-1.5}
\end{equation}
where $B$ is in G, and for $R$ between 1.02 and 10 $R_s$ approximately.

Using this model, we can predict the size of the radio sources at different heights in the corona, for a given height and size of the electron beam at its source in the corona.
The height of the source of the electron beam has been calculated between 19-77 Mm. The size of the beam transverse to the magnetic field is unknown in our case, so we examine several sizes between 1 and 36 Mm.
The results of the predictions are shown as red lines in figure \ref{vsheight}, using the lower limit for the acceleration site, 19 Mm.
As can be seen in this figure, even in the case of a large beam of 36 Mm, and in the case of a very low site of acceleration, the predicted size is significantly smaller than the observed size of the radio sources. 
Besides, the increase rate of the beam size is lower than the increase rate of the radio source with height.

\section{Discussion}
\label{sec:discussion}

\subsection{Geometry of the event}

\begin{figure}
\includegraphics[width=0.7\linewidth]{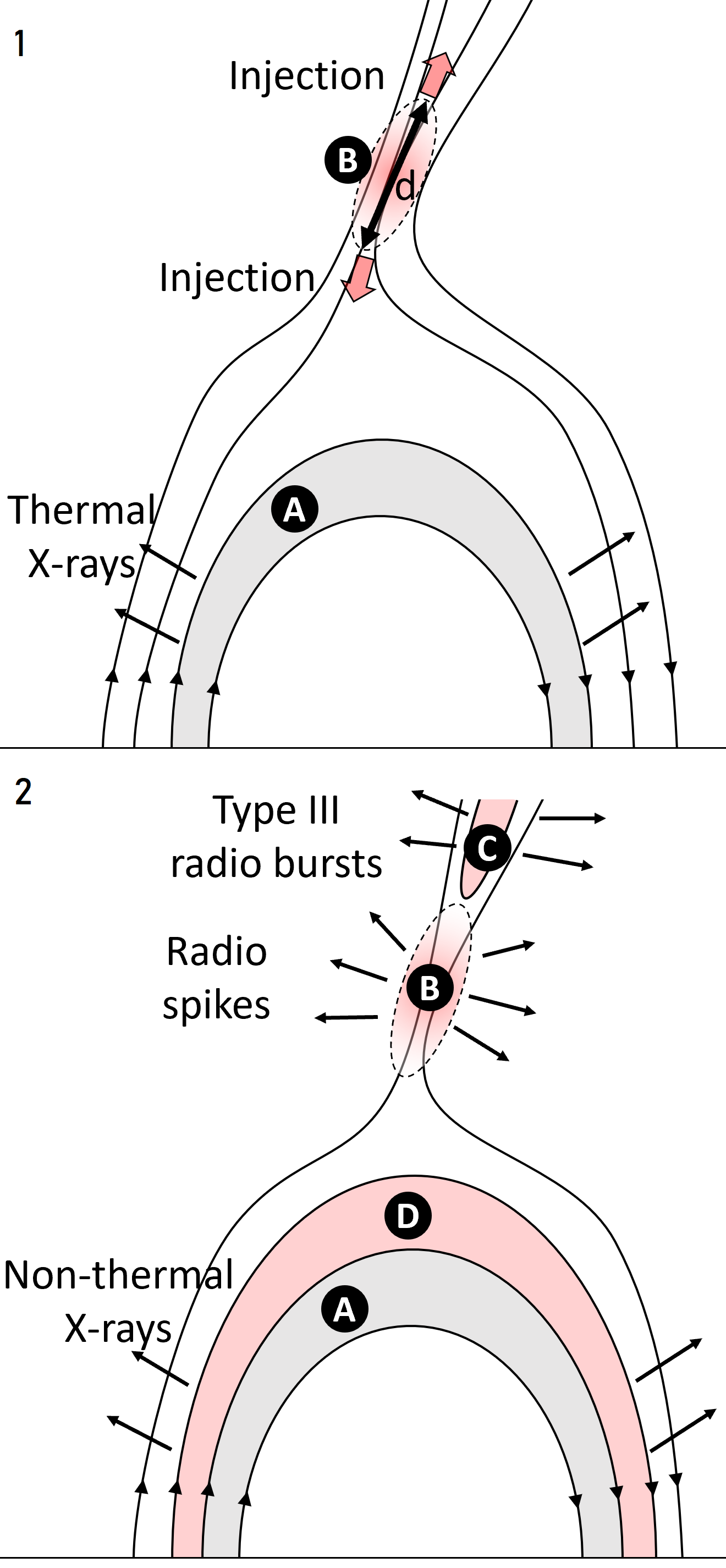}
\caption{Possible interpretation of the observations. Top: energy release occurs at a site of magnetic reconnection in the corona, above previously formed dense flare loops, and injecting energetic electrons both downwards in closed magnetic loops and upwards on open magnetic field lines. Bottom: this scenario results with radio emission in the form of radio spikes in the acceleration region, and type III radio bursts associated with the escaping beams of electrons, while X-ray are emitted by energetic electron precipitated in the lower, denser layers of the solar atmosphere.}
\label{schema}
\end{figure}

To interpret the observations presented in this paper, we make the following assumptions regarding the geometry of the event:
\begin{itemize}
\item The X-ray radiation is emitted in closed magnetic loops located below the site of energy release in the corona; the coronal X-ray sources are possibly emitted close to the site of energy release. This is a common assumption of flare models.
\item Type III radio bursts are produced by electron beams propagating along magnetic field lines open to the interplanetary medium. This is supported by the fact that the observed type III bursts extend from high frequencies consistent with coronal plasma densities, to low frequencies consistent with plasma densities of the interplanetary space. We assume that the origin of these electron beams is at or above the site of energy release. 
\end{itemize}

With these assumptions, heights (or densities) deduced from X-rays can be considered as lower (or higher) limits on the height (or density) of the region of energy release. Similarly, densities associated with the starting frequencies of type III radio bursts can be considered as lower limits on the density of the region of energy release.
Radio spikes have usually been associated with the site of energy release as they are emitted at a frequency slightly higher than type III radio bursts, as it is the case in this study.
The combination of hard X-ray and decimetric radio emissions provides therefore diagnostics to significantly constrain the density of the acceleration region in the corona. 
In the present event, we showed that the radio spikes have a temporal evolution similar to the excess non-thermal X-ray emission, supporting the idea that the process leading to radio spike emission has a close relationship with the process of electron acceleration itself.

Our interpretation of this event is summarized by the cartoons in figure \ref{schema}. We are focusing on an energy release event taking place during the gradual phase of a solar flare, and therefore pre-existing dense flare loops are present and emitting thermal X-rays (labeled A in the cartoon). 
Impulsive energy release happens creating the acceleration region (labeled B in the cartoon), where reconnection happens between closed and open magnetic field lines, and energetic electrons are accelerated and then injected downwards towards the flare loops, and upwards in the magnetic flux tube open to the heliosphere. The upward propagating electron beam is responsible for type III radio bursts (labeled C in the cartoon) while the downward propagating electrons are emitting X-ray bremsstrahlung emission when interacting with dense plasma (labeled D in the cartoon). Radio spikes are emitted close to the acceleration region.

\subsection{Relation between X-ray and radio emitting energetic electrons}

The event presented in this paper is an example of X-ray and radio emissions from energetic electrons particularly well correlated in time. A small increase of non-thermal X-ray emission corresponds to the observations of radio spikes in the 540-640 MHz and type III radio burst emission extending from hundreds of MHz to tens of kHz. This time correlation suggests a common origin for energetic electrons at the origin of both the X-ray and radio emissions. 
This is also supported by the comparison of different densities calculated from the observations: the lower limit of the plasma frequency in the X-ray emitting source is slightly higher than the higher frequencies at which spikes and type III radio bursts are detected, which is coherent with the geometry of the event assumed and described above.

We identified a correlation between fine temporal structures in the non-thermal X-ray flux and the radio spikes occurrence, which is coherent with the idea that spikes are emitted at or with a close relation to the acceleration region. 
Several studies reported that although radio decimetric spikes are usually associated with X-ray bursts, spikes and X-ray emissions are usually not closely correlated  \citep{benz_etal_2002,battaglia_benz_2009}. The present event is therefore of particular interest: in this case, time delays or structures induced during the injection may be reduced in comparison to events previously studied.
These radio spikes are emitted at a frequency of the same order as the type III radio bursts starting frequency or slightly higher, as it has been reported in many previous studies; moreover, \cite{benz_etal_2002} and \cite{battaglia_benz_2009} reported that spikes sources were usually emitted higher in the corona than the X-ray emission sites. This indicates that it is unlikely that spikes are emitted by X-ray emitting electrons. Rather, the time correlation on fine timescales between the X-ray flux and the radio spikes is likely linked to the acceleration process itself. The fragmentation of the spike emission could therefore be a signature of the fragmented nature of the acceleration process, for instance on a fragmented current sheet in the corona, as it was previously postulated by \cite{benz_1985}, or at fragmented plasmoids during a plasmoid merging process \citep{karlicky_etal_2013,karlicky_marian_2014}. 

Regarding the type III radio burst emission, a low-high-low evolution of the type III radio burst starting frequency is reported for the present event, similarly to the trends reported in \cite{reid_etal_2014}. As shown in previous studies, we observe a weak correlation between the starting frequency of the type III radio bursts and the non-thermal X-ray flux, which is another indication of the common origin of the X-ray emitting electrons and the electron beams responsible for the type III bursts. 

The present event, with its faint non-thermal X-ray signature, and the absence of detectable soft X-ray increase, is similar to the cases studied by \cite{james_etal_2017}, and the number of X-ray emitting electrons, $2 \times 10^{33}$ electrons$/$s, is comparable to the largest number of electrons in the events of their study. A similar number of electron would be enough to produce the type III radio burst emission, suggesting that similarly to the events presented in \cite{james_etal_2017}, and unlike the more intense events studied by \cite{lin_and_hudson_1971} and \cite{krucker_etal_2007}, the number of electrons accelerated upward could be more than a few percents of the number of electrons precipitated in the corona.


\subsection{Properties of the acceleration region}

By combining X-ray diagnostics of electrons with radio observations, and the relation between the electron spectral index, the type III radio burst starting frequency and the properties of the acceleration region of \cite{reid_etal_2014}, we derived an electron beam size in the range 0.5-11.4 Mm and the height of the acceleration region between 19 and 77 Mm, which is comparable to values found in many events by \cite{reid_etal_2014}. Note that the lower limit on the acceleration region height of 19 Mm is set by the height of the X-ray emission. This result confirms that the acceleration process occurs high in the corona.
The spectral index of the electron distribution is -4, which is on the side of harder spectra compared to the events presented in \cite{reid_etal_2014}. A harder spectra generally results in a smaller propagation distance of the electron beam before it becomes unstable to Langmuir waves: this is consistent with the fact that the higher starting frequencies observed in this event are close the frequencies of radio spikes, and consistent with a small beam size.

We note that our analysis provides an estimate on the elongation of electron beams, rather than an indication on the spatial scale for the electron acceleration itself. In the scenario where the electron cloud is accelerated and formed as a whole during the energization process, the elongation of this electron cloud could be assimilated to the size of the region of acceleration. However, the presence of spikes can be interpreted as a signature of numerous small-scale sites for the acceleration process, from which a continuous injection of the accelerated electrons will form the electron beam on a greater length than the acceleration region itself. This would be consistent with the conclusions from \cite{benz_1985}, where the analysis of radio spikes was used to estimate an acceleration spatial scale of 200 km, and from \cite{chen_etal_2018}, who used high-resolution imaging-spectroscopy of type III radio bursts close to the acceleration site in the corona to calculate an upper limit of 600 km$^2$ on the acceleration site.

\subsection{Evidence for radio-wave scattering in the high corona}

The radio source sizes in the 30-50 MHz range are in the range expected from previous observational studies, as it has been compiled by \cite{kontar_etal_2019}. 
These sizes and their radial evolution are compared to the expected radial evolution of the size of the electron beams at the origin of the type III radio bursts. 
Using the lower limit on the height of the acceleration site for the electron beam, assuming different beam size (perpendicular to the magnetic field) between 1 and 32 Mm, and
assuming that the expansion of the radio beam is controlled by the expansion of the magnetic field in the heliosphere, the expected sizes of the electron beams is calculated as a function of height in the corona.
If the coronal magnetic field follows typical models such as the one derived by \cite{dulk_coronal_1978}, the source size of the radio emission should be significantly smaller than what is observed. We note that by taking the lower limit on the height of the acceleration region, we produce an upper limit on the electron beam size at a given height in the corona.
We also showed that the source size growth rate with increasing radial distance from the Sun is higher than expected.
These observations provide further evidence that radio-wave scattering is a dominant factor in determining the radio source size in the heliosphere, as it was discussed in \cite{sharykin_etal_2018}. In this paper, the expected size of the radio-producing electron beam was derived from the combination of X-ray diagnostics in the low corona and radio observations at hundreds of MHz. This approach is completely different from the studies by \cite{kontar_nature_2017,sharykin_etal_2018} which used the fine structures in the LOFAR observations to estimate the expected sizes of the electron beams, and found similar scales for these sizes.

\section{Conclusion}

We analyzed in detail a group of type III burst emission that coincide with a small increase in X-ray during a short time interval during the gradual phase of the flare. 
The close relationship in time and space between the X-ray and radio emission suggests that the energetic electrons responsible of the type III burst emission have been accelerated during a reconnection event in the low corona, during the flare. A low-high-low trend of the starting frequency of the type III radio bursts has been observed, similar to what has been previously reported in other events.
The X-ray bursts and radio type III bursts are also associated with decimetric radio spikes in the 540-640 MHz frequency ranges, with a average duration of 0.14 seconds. 
We observed a correlation on fine time scales between the density of the radio spikes and the non-thermal X-ray excess flux, providing evidence that radio decimetric spikes are closely related the the source of accelerated electrons. 
Such a correlation on fine time scales between spikes and X-ray emissions is not commonly seen and provide further evidence that the radio spikes are a signature of the fragmentation of the mechanism of particle acceleration.
In this event, the non-thermal X-ray signature associated to the radio spikes and radio type III bursts remains very faint. 
The spectral index of non-thermal X-ray emitting energetic electrons is -4, and their number is about $2 \times 10{33}$ electrons/s, which is on the order of the number of electrons required to produce the type III radio bursts.
Using the spectral properties of the electrons deduced from the X-ray analysis and the starting frequencies of the type III radio bursts, 
we calculated that the density of the acceleration region is constrained between $1-5 \times 10^9$ cm$^{-3}$, which correspond, with a density model of the corona, to a height between 19 and 77 Mm for the site of acceleration. 
The elongation of the radio-producing electron beam at its source is in the range 0.5-7.8 Mm during the first X-ray burst and 4.1-11.4 Mm during the second.
Assuming that the size of the electron beams is governed by the expansion of the magnetic flux tubes on which they are injected, we calculated the expected radial evolution of the beam size, for a beam accelerated at a height of 19 Mm with a beam size between 1 and 32 Mm at its source.
The radio source sizes in the inner heliosphere were calculated using the spectro-imaging capability of the LOFAR radiotelescope in the 30-50 MHz range, and are wider than the expected size of the electron beam.
The evolution of the radio source size with frequency shows that the radio source extent increases beyond what is expected during the propagation of the electron beam and is not consistent with the only expansion of the magnetic field tube with increasing radial distance from the Sun. This study provide further evidence that radio-wave scattering on turbulent fluctuations of the ambient plasma density participate in a large proportion to the observed radio source size. 


\begin{acknowledgements}
We thank Bin Chen for a fruitful discussion on the characteristics of the acceleration sites in the corona.
We thank the LOFAR consortium, the RHESSI team, the AIA team and the GOES team for providing the data used in this paper.
We also thank the RHESSI and AIA team for the software provided to analyze the data in the SolarSoft suite.
We thank the RSDB service at LESIA / USN (Observatoire de Paris) for making the ORFEES data available.
We equally thanks the radio monitoring service at LESIA (Observatoire de Paris) to provide value-added data that have been used for this study.
The authors acknowledge the Nançay Radio Observatory / Unité Scientifique de Nançay of the Observatoire de Paris (USR 704-CNRS, supported by Université d’Orléans, OSUC, and Région Centre in France) for providing access to NDA observations accessible online at \url{http://www.obs-nancay.fr}.
We also acknowledge support from the International Space Science Institute for the LOFAR \url{http://www.issibern.ch/teams/lofar/} and solar flare \url{http://www.issibern.ch/teams/solflareconnectsolenerg/teams}.
\end{acknowledgements}

\bibliographystyle{aa} 
\bibliography{zmabiblio} 


\appendix
\section{Appendix: Measurements of LOFAR source sizes and positions}

LOFAR observations consists of radio flux measurements $F_i = F_i(x_i,y_i)$ with $i$ between 1 and $N_b$, the number of beams, where $(x_i,y_i)$ are the positions of each beam.

\subsection{Contours of the source}
\label{app:contours}

The LOFAR image can be reconstructed by interpolating the intensities measured at the different beam positions $(x_i,y_i)$  to provide a dirty image $I(x,y)$. This image is the convolution of the radio source with the telescope point spread function, which include side lobes in the image. Nevertheless, when side lobes can be neglected, the 50 \% contour of the obtained dirty intensity map is a good approximation of the radio source. 
The centroid position and the source size are determined as described in \cite{kontar_etal_2019}:
the centroid position $(\overline{x},\overline{y})$ is calculated as the first normalized moments of the intensity distribution within the contour:

\begin{equation}
\overline{x} = \frac{ \int^{\infty}_{-\infty} x I(x,y) dx dy }{\int^{\infty}_{-\infty} I(x,y) dx dy}, \ \overline{y} = \frac{ \int^{\infty}_{-\infty} y I(x,y) dx dy }{\int^{\infty}_{-\infty} I(x,y) dx dy}
\end{equation}
The size of the source is calculated using the second normalized moments (variances) of the distribution  $(\sigma_x,\sigma_y)$:
\begin{equation}
\sigma_x^2 = \frac{ \int^{\infty}_{-\infty} (x-\overline{x})^2 I(x,y) dx dy }{\int^{\infty}_{-\infty} I(x,y) dx dy}, \
\sigma_y^2 = \frac{ \int^{\infty}_{-\infty} (y-\overline{y})^2 I(x,y) dx dy }{\int^{\infty}_{-\infty} I(x,y) dx dy}
\end{equation}
The size of the radio source is then characterized as the full width half maximum ($FWHM$) such as:
\begin{equation}
FWHM_{x,y} = 2 \sqrt{2 \ln 2} \ \sigma_{x,y}
\end{equation}

The associated statistical errors for the centroid positions are:
\begin{equation}
\delta \overline{x} \approx \frac{\sigma_x}{\sqrt{N}}, \delta \overline{y} \approx \frac{\sigma_y}{\sqrt{N}}
\end{equation}
where $N$ is the number of photons used to construct the image, and the uncertainty on the sizes are:
\begin{equation}
\delta FWHM_{x,y} \approx 2 \sqrt{2 \ln 2} \frac{\sigma_{x,y}}{\sqrt{2N}}
\end{equation}

\subsection{Elliptical Gaussian fit to the data}
\label{app:fit}

This method is adapted to the procedure described in \cite{kontar_nature_2017}. 
Assuming that the radio sources $S$ have the shape of an elliptical Gaussian:
\begin{equation}
S(x,y) = S_0 \exp{\left( -\frac{ x'^2 }{ 2 \sigma_x^2} -  \frac{ y'^2 }{ 2 \sigma_y^2} \right)} + B_0
\end{equation}
where $x' = \left(x-x_0\right) \cos{\left(T\right)} - \left(y-y_0\right) \sin{\left(T\right)}$ and $y' = \left(x-x_0\right) \sin{\left(T\right)} - \left(y+y_0\right) \cos{\left(T\right)}$, $T$ being the rotation angle from the $x$ axis, and $B_0$ being a constant background.

We fit this model to the measured flux by minimizing the $\chi^2$:
\begin{equation}
\chi^2 = \sum_{i=1}^{N_b} \frac{ \left( F_i - S \left( x_i,y_i; S_0,x_0,y_0,\sigma_x,\sigma_y,T,B_0 \right) \right)^2 }{\delta F^2}
\end{equation}
where $\delta F$ is the uncertainty on the measured flux. 

The uncertainties on the source position and size are calculated using:
\begin{equation}
\delta x_0 \approx \sqrt{\frac{2}{\pi}} \frac{\sigma_x \delta F}{\sigma_y S_0} h, \ \delta y_0 \approx \sqrt{\frac{2}{\pi}} \frac{\sigma_y \delta F}{\sigma_x S_0} h
\end{equation}
where $h$ is the spatial resolution.

\end{document}